\documentclass[lettersize,journal]{IEEEtran}
\usepackage{amsmath,amsfonts}
\usepackage{algorithmic}
\usepackage{algorithm}
\usepackage{array}
\usepackage[caption=false,font=normalsize,labelfont=rm,textfont=rm]{subfig}

\usepackage{textcomp}
\usepackage{bbding}
\usepackage{stfloats}
\usepackage[colorlinks,
linkcolor = blue, 
anchorcolor=blue, 
citecolor=blue, 
urlcolor=blue,
]{hyperref}
\usepackage[english]{babel}
\usepackage{verbatim}
\usepackage{graphicx}
\usepackage{color}
\usepackage{cite}
\newcommand{\comments}[1]{}
\usepackage{booktabs}
\usepackage{multirow}
\usepackage{makecell}
\newcommand{\METHODNAME}[1]{\(\text{WavInWav}\)}
\newcommand{\MODELNAME}[1]{\(\text{WavINN}\)}
\begin{document}

\title{WavInWav: Time-domain Speech Hiding via Invertible Neural Network}
\author{Wei Fan, Kejiang Chen, Xiangkun Wang, Weiming Zhang, Nenghai Yu
\thanks{This work was supported in part by the Natural Science Foundation of China under Grant 62102386, 62002334, 62072421 and 62121002.}%
\thanks{All the authors are with CAS Key Laboratory of Electromagnetic Space Information, School of Information Science and Technology, University of Science
and Technology of China, Hefei 230026, China.}
\thanks{Corresponding authors: Kejiang Chen and Weiming Zhang (Email:chenkj@ustc.edu.cn, zhangwm@ustc.edu.cn).}
}

\markboth{Under Review}%
{Shell \MakeLowercase{\textit{et al.}}: A Sample Article Using IEEEtran.cls for IEEE Journals}


\maketitle

\begin{abstract}
Data hiding is essential for secure communication across digital media, and recent advances in Deep Neural Networks (DNNs) provide enhanced methods for embedding secret information effectively. However, previous audio hiding methods often result in unsatisfactory quality when recovering secret audio, due to their inherent limitations in the modeling of time-frequency relationships. In this paper, we explore these limitations and introduce a new DNN-based approach. We use a flow-based invertible neural network to establish a direct link between stego audio, cover audio, and secret audio, enhancing the reversibility of embedding and extracting messages. To address common issues from time-frequency transformations that degrade secret audio quality during recovery, we implement a time-frequency loss on the time-domain signal. This approach not only retains the benefits of time-frequency constraints but also enhances the reversibility of message recovery, which is vital for practical applications.  We also add an encryption technique to protect the hidden data from unauthorized access.  Experimental results on the VCTK and LibriSpeech datasets demonstrate that our method outperforms previous approaches in terms of subjective and objective metrics and exhibits robustness to various types of noise, suggesting its utility in targeted secure communication scenarios.

\end{abstract}

\comments{keywords: TODO}
\section{Introduction}
\IEEEPARstart{D}{ata hiding} is the science and technology of embedding secret information into cover object without perceptibly altering it. During the hiding process, a secret message is concealed within a cover message in an imperceptible manner. During the revealing process, the secret message is extracted from the stego message. With the increasing popularity of digital media, such as digital audio~\cite{chen2021distribution,cuiicassp2020imginaudio,kreuk2019hide}, images~\cite{baluja2017hiding,zhu2018hidden}, and videos~\cite{yang2019video}, these media forms are widely employed as cover objects in data hiding applications.

For digital audio, numerous internet communication applications (e.g., iTunes, YouTube, Twitter) provide audio storage and communication services, providing a diverse range of digital audio cover objects. Furthermore, leading music-sharing platforms such as Spotify\footnote{\url{https://open.spotify.com/}}, SoundCloud\footnote{\url{https://soundcloud.com/}}, and Bandcamp\footnote{\url{https://bandcamp.com/}} offer additional opportunities for implementing audio data hiding techniques. These platforms collectively broaden the spectrum of available channels for audio data hiding, creating an environment conducive to the secure transmission of hidden messages under the guise of sharing audio.
\begin{figure}
    \centering
    \subfloat[Traditional framework relying on spectrums.]{
        \label{fig:inn_a}
        \includegraphics[width=0.9\linewidth]{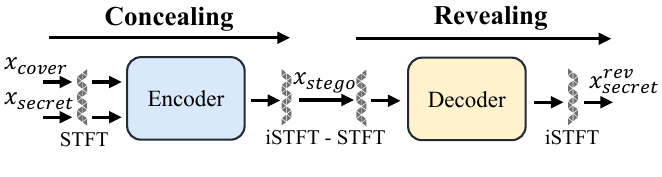}
}\hfill
    \subfloat[Our flow-based framework with time-domain hiding.]{
        \label{fig:inn_b}
        \includegraphics[width=0.9\linewidth]{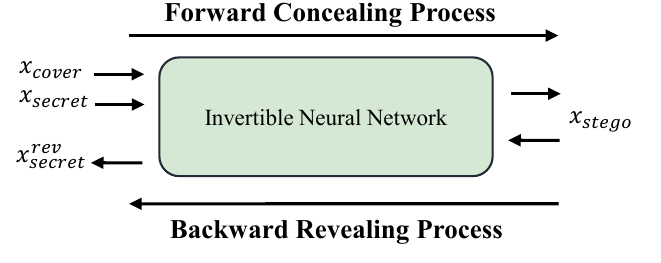}
}\hfill
    
    \caption{Comparison between framework relying on spectrums and our flow-based framework with time-domain hiding.}
    \label{fig:inn}
\end{figure}
Traditional data hiding techniques in audio often involve manually modifying audio data using rules such as altering the least significant bit (LSB) of the audio~\cite{yan2008reversible, lsbascii, lsbrsa, nassrullah2020enhancement}. The domains involved in these traditional methods encompass not only the commonly used time domain but also various transform domains and compressed domains~\cite{fft2009fallahpour, hongson2013dct, dwt2002cvejic, mp32016bazyar, ren2021silk, ren2021aac, tian2015voip, mazurczyk2008voip}. However, these traditional methods have limited capacity (for instance, LSB methods typically use only one or two bits per sample), making them unsuitable for concealing higher information payloads such as images or audio~\cite{yang2019video}. Recently, researchers have employed neural networks as the function for audio data hiding, enabling the concealment of multimedia data within audio. \IEEEpubidadjcol 
Given the sensitivity of the Human Auditory System (HAS) to audio frequencies, modeling time-frequency relationship of the audio data becomes crucial. Therefore, references~\cite{cuiicassp2020imginaudio, geleta2022pixinwav, ros2023pixinwav2} hide image data within the time-frequency-domain data obtained after performing Short-Time Fourier Transform (STFT) on the audio data. STFT transforms the audio into a complex matrix consisting of the Fourier transforms of different time frames, and its inverse transform reconstructs the time-domain waveform from the complex matrix. Since STFT transformation introduces distortion, Kreuk~\textit{et al.}~\cite{kreuk2019hide} incorporate differentiable STFT layers into the encoder and decoder to facilitate the learning of embedding and extraction processes in the presence of communication channel distortion. Their approach effectively conceals one audio segment within another.

However, previous works on hiding audio within audio primarily focus on the quality of the stego audio while giving less consideration to the quality of the recovered secret audio. Specifically, they recover the amplitude spectrum of the secret audio from the stego audio and reconstruct the secret audio relying on the inferred amplitude spectrum. But this reconstruction process is far from perfect, as the amplitude spectrum of the secret audio recovered from a distorted channel is invalid. In reality, there does not exist a reasonable audio with the same invalid amplitude spectrum, leading to a lower quality of the reconstructed secret audio.

Considering the advantages of directly using speech audio as the secret message, such as preserving speaker characteristics, intonation, and rhythm for subsequent recognition and authentication, the quality of the message audio is crucial for audio hiding. However, the low-quality secret audio from previous methods undermines the advantages of concealing speech audio. To address the issue of previous methods relying on distorted spectrums during secret reconstruction, we propose directly hiding the message in the time domain while utilizing time-frequency constraints to preserve the advantages of modeling the time-frequency relationship, while avoiding channel distortion caused by time-frequency transformations. Specifically, we employ a flow-based invertible network to jointly model the hiding and revealing processes, which further enhances the reversibility of message embedding and extraction compared to encoder-decoder architectures. The distinction between the proposed method and previous approaches is illustrated in Fig.~\ref{fig:inn}. 

Building upon this framework, and in response to the challenges posed by lossy channels,  we introduce a noise layer featuring a range of common differentiable noises during our training phase. This enhancement allows the model to efficiently process the extraction of secret audio even in suboptimal channel conditions, ensuring consistently high-quality recovery across various scenarios. Moreover, to safeguard the secret audio from unauthorized access, we incorporate an encryption-decryption module within our system. Secret audio can only be decrypted with a matching decryption key, significantly enhancing system security. Experiments, both subjective and objective, conducted on the VCTK and LibriSpeech datasets demonstrate the significant superiority of our method in terms of stego and recovered secret quality compared to previous approaches.  Furthermore, our method exhibits robustness against various common distortions.

Our contributions can be summarized as follows:
\begin{itemize}
    
    \item To the best of our knowledge, we are the first to explore the hiding and subsequently end-to-end recovering of a complete secret audio segment within a cover audio. Unlike previous methods that reconstructed secret audio from the spectrum, we model the hiding and revealing processes directly in the time domain, using time-frequency loss as a constraint to avoid distortion during time-frequency transformation, thus enhancing the quality of the secret audio.

    \item We employ a novel invertible network that simultaneously trains the hiding and revealing processes. Compared to traditional encoder-decoder architectures, our introduced invertible network exhibits better reversibility in the message embedding and extraction processes.
    
    \item We use a carefully designed composite noise layer to enhance the robustness of the invertible network. Additionally, to protect the secret messages from unauthorized extraction, we introduce a simple yet effective encryption-decryption module, enhancing the overall security of the system.

\end{itemize}
The remaining sections of this paper are organized as follows. Section~\ref{sec:related-work} conducts a review of related works. Section~\ref{sec:method} introduces our proposed method in detail. Section~\ref{sec:experiment} gives the experimental results and discussions, while Section~\ref{sec:conclusion} serves as the conclusion to our work.

\section{Related Works}
\label{sec:related-work}
\subsection{Traditional Audio Data Hiding}
Traditionally, audio data hiding leverage the real or perceive redundancies within cover signals, employing techniques across various domains like time domain, transform domain, and compressed domain.

In the time domain, one of the earliest data hiding methods embed secret information into least significant bits (LSB) of each cover sample. References~\cite{sridevi2009efficient,lsbrsa,lsbascii, yan2008reversible} encrypt the secret into ciphertext and utilize the LSB method to embed it into cover audio to enhance security. Nassrullah~\textit{et al.}~\cite{nassrullah2020enhancement} propose an adaptive audio data hiding strategy based on LSB, dynamically balancing distortion rates and embedding capacity as required. Chen~\textit{et al.}~\cite{chen2023imperceptible} employ psychoacoustic models alongside adversarial examples, enhancing both the imperceptibility and undetectability of traditional LSB techniques. These LSB methods offer relatively high hiding capacities but are vulnearable to modification. Furthermore, for multimedia data such as images or audio, the hiding capacity of LSB (one or two bits per sample) remains limited.  Other time-domain data hiding methods include phase coding, echo hiding, and spreading, considering the characteristics of the HAS to hide information in areas imperceptible to humans. Djebbar~\textit{et al.}~\cite{djebbar2013phase} use the low sensitivity of HAS to phase distortion to hide information by altering audio phase values, providing a degree of robustness. Oh~\textit{et al.}~\cite{echo2001}  utilize the temporal masking effect of HAS, introducing slightly delayed echo signals into cover audio, rendering them robust against common signal processing attacks. Matsuoka~\textit{et al.}~\cite{spread2006} expand narrowband secret information across wider frequency ranges for information hiding. These non-LSB techniques are more robust to modifications and compression but offer lower capacities. 

Beyond the time domain, discrete cosine transform (DCT)~\cite{hongson2013dct}, fast Fourier transform (FFT)~\cite{fft2009fallahpour}, and discrete wavelet transform (DWT)~\cite{dwt2002cvejic} are employed in audio data hiding. Additionally, considering audio compression during transmission, particularly through MP3 and VoIP, references~\cite{mp32016bazyar, mazurczyk2008voip, ren2021silk, tian2015voip, ren2021aac} directly embed within the compressed domain to alleviate potential losses incurred during the compression process.

In general, the capacity of traditional audio data hiding methods is far from sufficient for multimedia data. Differing from traditional methods, we employ deep neural networks to achieve high perceptual transparency and increased hiding capacity while maintaining a certain level of robustness.

\subsection{Deep-learning-based Audio Data Hiding}
In recent years, several deep-learning-based data hiding schemes have been proposed. Initially, there was a proposal to train neural networks for hiding images within other images~\cite{baluja2017hiding}. In this approach, an encoder network is utilized to conceal a secret image within a cover image to generate a stego image, while a decoder network is employed to recover the secret image from the stego image. Zhu~\textit{et al.}~\cite{zhu2018hidden} extend this work by introducing adversarial loss terms and noise layers, enhancing robustness against various forms of distortion.

Given that the HAS is more sensitive than the Human Visual System (HVS)~\cite{cvejic2004algorithms}, digital audio data hiding is typically more challenging than image hiding.  Kreuk~\textit{et al.}~\cite{kreuk2019hide} point out that steganographic models designed for visual content might not be suitable for audio. They suggest integrating differentiable STFT layers between the encoder and decoder to adapt the spectrogram of audio to the framework of image hiding. While this method effectively conceals and recovers identifiable secret audio, the quality of the recovered audio is compromised. Extending beyond audio, images and videos can also serve as hidden information for audio data hiding. Yang~\textit{et al.}\cite{yang2019video} suggest using the reversibility of flow-based models to hide video frames within audio. References~\cite{cuiicassp2020imginaudio, geleta2022pixinwav, ros2023pixinwav2} conceal images within the audio spectrogram. Chen~\textit{et al.}~\cite{chen2021distribution} introduce a provably secure steganographic method maintaining distribution using audio generation models such as WaveGlow~\cite{prenger2019waveglow} and WaveNet~\cite{oord2016wavenet}. Furthermore, Takahashi~\textit{et al.}~\cite{takahashi2021source} propose a robust audio hiding method for instrument source separation, concealing byte-level information within different musical instrument tracks. Previous studies have demonstrated the immense potential of DNNs in audio data hiding, however, their performance in extracting audio-type secret information remains unsatisfactory.

\subsection{Normalizing Flow-based Model}
Normalizing flow-based models exhibit a strong ability to directly capture complex probability distributions from data. The structure of a normalizing flow-based model is reversible, allowing it to efficiently perform both forward and inverse mappings using the same network and parameters. The property of reversible mappings makes the model particularly suitable for implementations as an Invertible Neural Network (INN). For instance, given a variable $x$ and a forward function $y = f_\theta(x)$, $x$ can be recovered through $x = f^{-1}_{\theta}(y)$, where $f^{-1}_{\theta}$ and $f_{\theta}$ share the parameters $\theta$. 

Normalizing flow-based models were initially introduced for tasks involving image generation. Pioneering work by Dinh~\textit{et al.}~\cite{dinh2014nice} first demonstrate the strong generative capabilities of normalizing flow-based models. Subsequently, Kingma~\textit{et al.}~\cite{kingma2018glow} enhance their utility in realistic image synthesis and manipulation by introducing invertible $1 \times 1$ convolutions in Glow model. Owing to their exceptional performance, flow-based models are applied in various image-related tasks. For example,  Ardizzone~\textit{et al.}~\cite{ardizzone2019guided} introduce conditional flow-based models for guiding image generation and coloring, addressing the task of natural image generation guided by a conditioning input. Xiao~\textit{et al.}~\cite{xiao2020rescale} leverage the bidirectional transformation capabilities of these models  for image super-resolution, mapping between low and high-resolution images. 

In the specialized area of data hiding, flow-based models  demonstrate considerable efficacy. Jing~\textit{et al.}~\cite{jing2021hinet} incorporate flow-based models into image hiding, exploring the potential of modeling image revealing as the inverse process of image concealing within an invertible network architecture. Following this, flow-based models are increasingly adopted in the field, with numerous studies affirming their superiority in handling large-capacity concealment tasks~\cite{guan2022deepmih, ren2022color}. Furthermore, Xu~\textit{et al.}~\cite{xu2022robust} introduce the conditional flow and noise layer into the invertible framework, thereby achieving significant improvements in robustness. Ma~\textit{et al.}~\cite{ma2022cin} propose an image watermarking network that combines invertible and non-invertible frameworks, utilizing the non-invertible part for handling quantization noise, effectively balancing the fidelity and robustness of the watermark. These enhancements in robustness showcase the resilience of flow-based models in concealing tasks. Overall, the success of these diverse implementations indicates the broad potential of flow-based models for applications requiring precise, bidirectional transformations.

While INNs have found numerous applications in image processing tasks, to the best of our knowledge, they have not been explored in the context of audio hiding. Given the reversible nature of flow-based models, which allows for precise, bidirectional transformations, this paper presents our initial attempt to model audio concealing and revealing as a pair of inverse processes using normalizing flow-based models.

\section{The Proposed Method}
\label{sec:method}
This section presents the proposed novel audio data hiding approach referred to as ``\METHODNAME~'', which involves hiding secret audio within a high-capacity and imperceptible cover audio. The core concept behind this approach lies in utilizing an invertible network to concurrently model the concealing and revealing processes of time-domain data while employing time-frequency constraints as training constraints. Additionally, to enhance the robustness of the proposed method against channel distortions, we incorporate a composite noise layer during the training process to simulate distortions. To further secure the hidden data against unauthorized extraction, we also integrate an encryption mechanism into our system. We outline the entire system architecture in Section~\ref{sec:method-overview}, explain the concept of time-domain hiding in Section~\ref{sec:method-Time-domain}, introduce the network architecture we employ in Section~\ref{sec:method-network}. Following this, Section~\ref{sec:method-security} is dedicated to detailing the implemented security mechanisms, including the specifics of the encryption and decryption processes.  Lastly, we present our chosen loss functions in Section~\ref{sec:method-loss}.

\subsection{Overview}
\renewcommand{\arraystretch}{1.25}
\begin{table}[htbp]
    \centering
    \caption{Notations Used in the \METHODNAME~ Framework}
    \label{tab:notations}
    \begin{tabular}{cl}
        \toprule
        \textbf{Notation} & \textbf{Description} \\
        \hline
        \( x^{orig}_{secret} \) & Original secret audio before encryption \\
        \( x_{secret} \) & Encrypted secret audio \\
        \( x_{cover} \) & Cover audio used to conceal the secret \\
        \( x_{stego} \) & Stego audio containing the encrypted secret audio \\
        \( x^{rev}_{secret} \) & Recovered secret from the stego audio \\
        \( x^{rev}_{cover} \) & Recovered cover audio from the stego audio \\
        \( x^{restored}_{secret} \) & Decrypted secret audio reconstructed from \( x^{rev}_{secret} \) \\
        \(r\) & Discarded information in the concealing process \\
        \(z\) & Auxiliary variable used in the revealing process\\
        \bottomrule
    \end{tabular}
\end{table}
\renewcommand{\arraystretch}{1} 

\begin{figure*}
    \centering
    \includegraphics[width=\linewidth]{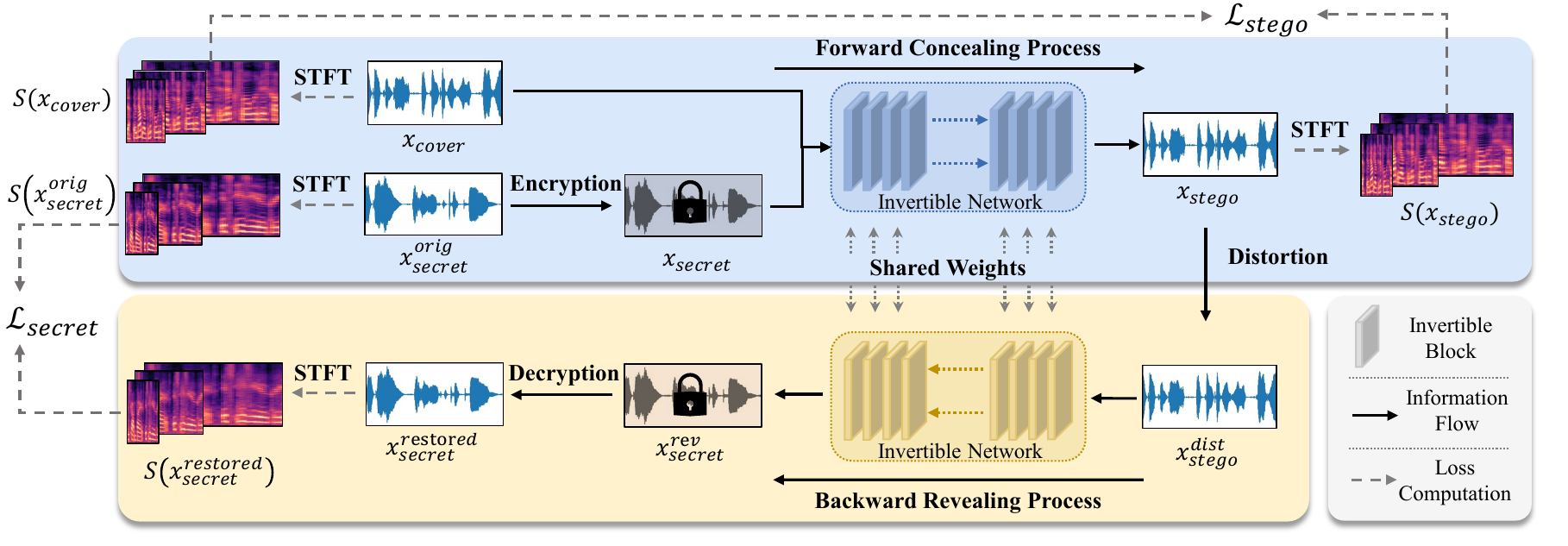}

    \caption{Overview of the proposed audio hiding method. The audio is hidden in the time domain, while loss is computed in the time-frequency domain.}
    \label{fig:time-domain}
\end{figure*}
\label{sec:method-overview}
Fig.~\ref{fig:time-domain} illustrates the overall framework of the proposed \METHODNAME~, which utilizes a flow-based invertible neural network, referred to as \MODELNAME~. In the concealing phase, the secret audio \( x^{orig}_{secret} \) undergoes encryption to form \( x_{secret} \). Subsequently, both the cover audio \( x_{cover} \) and the encrypted secret audio \( x_{secret} \) are input into \MODELNAME~ to produce \( x_{stego} \). This \( x_{stego} \) maintains perceptual similarity to \( x_{cover} \) while encapsulating \( x_{secret} \) within.

The reveal process utilizes the same network architecture as the concealment phase, with an identical parameter set, but reverses the information flow. During this backward phase, \MODELNAME~ processes the stego audio \( x_{stego} \) to extract the recovered secret \( x^{rev}_{secret} \) and the cover audio \( x^{rev}_{cover} \). Subsequently, the decrypted secret audio \( x^{restored}_{secret} \) is reconstructed from \( x^{rev}_{secret} \), completing the secure recovery process. Table~\ref{tab:notations} lists the notations used in this paper.

\subsection{Time-domain Audio Hiding}

\label{sec:method-Time-domain}
In this section, we provide a detailed explanation of the concept of hiding messages in the time domain. Spectrograms offer an intuitive representation of the frequency spectrum of an audio signal, which is more aligned with human auditory perception compared to the time-domain representation of audio signals. Therefore, previous methods often operate in the frequency domain, requiring the transformation of audio data into spectral data using the STFT for concealing in the frequency domain and subsequent inverse Short-Time Fourier Transform (iSTFT) for conversion back to the time domain. Specifically, for the cover audio $x_{cover}$ and the secret audio $x_{secret}$, we denote the frequency-domain hiding function as $E_f(\cdot)$ and the function for extracting secret messages from the frequency domain as $D_f(\cdot)$, with STFT denoted as $S(\cdot)$ and iSTFT denoted as $S^{-1}(\cdot)$. The forward concealment process in the frequency domain can be expressed as:
\begin{equation}
    S^{-1}\left(E_f\left(S\left(x_{cover}\right), S\left(x_{secret}\right)\right)\right) = x_{stego}.
\end{equation}

Subsequently, the process of extracting the secret audio from the stego audio can be represented as:
\begin{equation}
    S^{-1}\left(D_f\left(S\left(x_{stego}\right)\right)\right) = x^{rev}_{secret}.
    \label{eq:reveal}
\end{equation}

This approach is advantageous in maintaining perceptual consistency between the cover audio and the stego audio. However, it falls short in delivering satisfactory secret audio quality due to two inherent flaws. First, because the human auditory system is more sensitive to the amplitude spectrum of audio and phase spectra are often challenging to learn, frequency-domain audio hiding typically operates on amplitude spectra, leading to the loss of the original phase of the secret audio during the hiding process. The recovery of the secret audio phase must be inferred from the amplitude spectrum $D_f(S(x_{stego}))$. Furthermore, the STFT-iSTFT transformation process is lossy for the stego audio, meaning that $S^{-1}(S(x_{stego})) \neq x_{stego}$. Consequently, the amplitude spectrum $D_f(S(x_{stego}))$ of the recovered secret audio is also lossy.  This indicates that the recovered amplitude spectrum $D_f(S(x_{stego}))$ might not align with any realistic audio signal $x^{rev}_{secret}$ that satisfies the criteria of  Equation~\ref{eq:reveal}. In other words, a realistic audio that matches $D_f(S(x_{stego}))$ may not actually exist. This further contributes to the error between $x^{rev}_{secret}$ and $x_{secret}$, rendering the previous method unsatisfactory in terms of secret audio quality.

To address these two shortcomings and achieve the recovery of high-quality secret audio, we consider directly concealing information in the time domain. Fig.~\ref{fig:time-domain} demonstrates our concept of hiding the message in the time domain. We denote the time-domain hiding function as $E_t(\cdot)$ and the function for extracting secret messages in the time domain as $D_t(\cdot)$. The forward concealment process in the time domain can be expressed as:
\begin{equation}
    E_t\left(x_{cover}, x_{secret}\right) = x_{stego},
\end{equation}
where $x_{stego}$ represents the stego audio. The process of recovering the secret audio from the stego audio can be represented as:
\begin{equation}
    D_t\left(x_{stego}\right) = x^{rev}_{secret},
\end{equation}
where $x^{rev}_{secret}$ denotes the revealed secret.

By directly concealing information in the time domain, we circumvent the need for STFT during audio hiding and extraction, thus avoiding the impact of the lossy STFT-iSTFT transformation and the challenge of recovering audio from unrealistic spectrograms.

\subsection{Network Structure}
\label{sec:method-network}
\begin{figure*}
    \centering
    \includegraphics[width=0.95\linewidth]{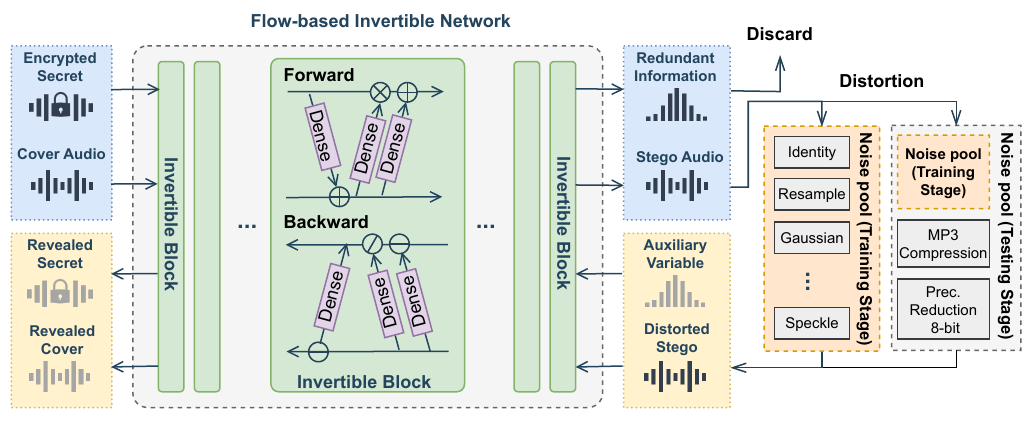}
    \caption{The structure of the invertible neural network and noise layers used in our approach. The forward and backward processes share the same parameters.}
    \label{fig:inv}
\end{figure*}

In this section, we provide a detailed overview of the architecture of the invertible neural network \MODELNAME~ that we employ. Fig.~\ref{fig:inv} illustrates the structure of the invertible neural network used in our approach. By using an invertible network, encoded features can be highly consistent with the features required by the decoder. This approach helps to limit the inclusion of redundant features in the encoding process, resulting in a more efficient and targeted representation of the necessary information. Specifically, \MODELNAME~ consists of a series of entirely invertible network modules, and its forward concealment process and backward reveal process are inverse operations. Thus, we denote the forward function as $f_{\theta}$, and the backward function can be represented with the same parameters $\theta$ as $f^{-1}_{\theta}$.

In the forward process, $f_{\theta}$ takes a pair of cover audio $x_{cover} \in \mathbb{R}^{1 \times t}$ and encrypted secret $x_{secret}$ as input and then outputs the stego audio $x_{stego}$ along with redundant information $r$. For the backward decoding process, $f^{-1}_{\theta}$ takes the stego audio $x_{stego}$ and an auxiliary variable $z$ as input. After passing through a series of network modules that share parameters with the forward network, it recovers the secret $x^{rev}_{secret}$. It should be noted that the redundant information $r$ is discarded during this process. Therefore, it is necessary to resample an auxiliary variable $z$ from a Gaussian distribution to replace the redundant information $r$ as part of the input during the decoding process.

\subsubsection{Invertible Blocks}

\begin{figure}
    \centering
    \includegraphics[width=\linewidth]{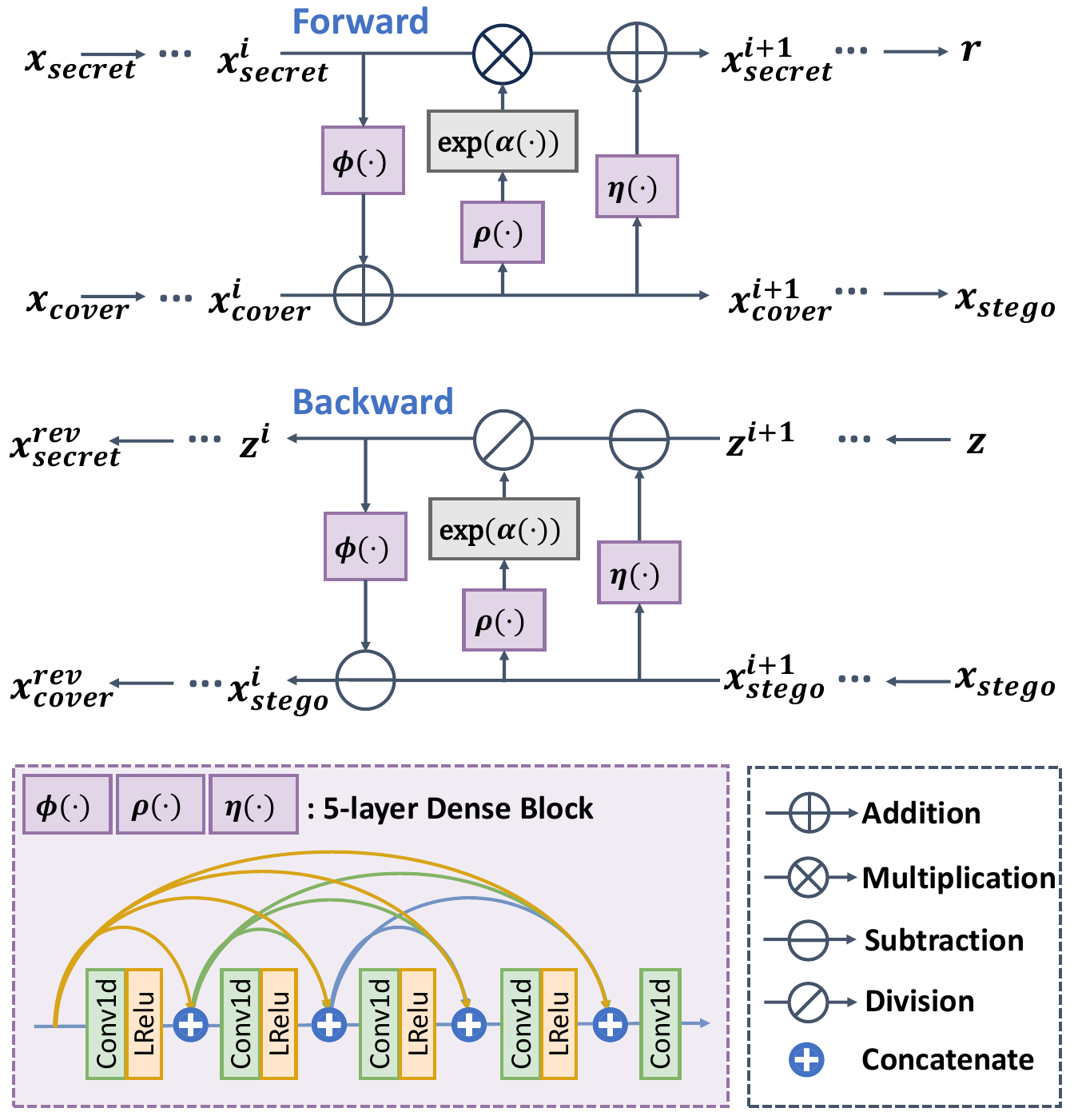}
    \caption{The backbone of the $i$-th invertible block used in our approach.}
    \label{fig:block}
\end{figure}

As illustrated in Fig.~\ref{fig:block}, the concealment network and the reveal network share submodules and network parameters, but the flow of information is in opposite directions. Both the concealment network and the reveal network consist of $M$ concealment blocks.
For the $i$-th concealment block in the forward process, given inputs $x^{i}_{cover}$ and $x^{i}_{secret}$, the outputs $x^{i+1}_{cover}$ and $x^{i+1}_{secret}$ can be expressed as follows:
\begin{equation}
    x^{i+1}_{cover} = x^i_{cover} + \phi\left(x^i_{secret}\right),
\end{equation}
\begin{equation}
x^{i+1}_{secret} = x^{i}_{secret} \cdot \exp\left(\alpha\left(\rho\left(x^{i+1}_{cover}\right)\right)\right) + \eta\left(x^{i+1}_{cover}\right),
\end{equation}
where $\alpha(\cdot)$ is a sigmoid function scaled by a constant factor used as a clamp, $\cdot$ denotes the dot product operation, and $\rho(\cdot)$, $\eta(\cdot)$, and $\phi(\cdot)$ can be arbitrary functions, not necessarily invertible ones. In this context, we employ one-dimensional dense blocks as the functions $\rho(\cdot)$, $\eta(\cdot)$, and $\phi(\cdot)$ due to their efficacy in audio concealment tasks. Dense blocks are highly effective at promoting the flow of information across layers, enhancing the reuse of features, and optimizing the utilization of parameters. These factors are essential for the effective concealment of audio.

For a network with $M$ concealment blocks, during the forward process, the first concealment block takes inputs $x_{cover}$ and $x_{secret}$, while the $M$-th block, which is the last block, takes inputs $x^{M}_{cover}$ and $x^{M}_{secret}$. The outputs are the final stego audio $x_{stego}$ and redundant information $r$.

In the backward process for the $i$-th concealment block, the information flow is reversed compared to the forward process, going from the $(i+1)$-th block to the $i$-th block. Therefore, the inputs are $x^{i+1}_{stego}$ and $z^{i+1}$, and the outputs $x^{i}_{stego}$ and $z^{i}$ can be expressed as follows:
\begin{equation}
    z^i = \left(z^{i+1} - \eta\left(x^i_{stego}\right)\right) \cdot \exp\left(-\alpha\left(\rho\left(x^{i+1}_{stego}\right)\right)\right),
\end{equation}
\begin{equation}
    x^{i}_{stego} = x^{i+1}_{stego} - \phi\left(z^{i}\right).
\end{equation}

For a network with $M$ concealment blocks, during the backward process, the $M$-th block takes inputs $x_{stego}$ and $z$, and the outputs are $x^{M-1}_{stego}$ and $z^{M-1}$, while the outputs of the first block are the ultimately revealed cover audio $x^{rev}_{cover}$ and revealed secret $x^{rev}_{secret}$.

\label{sec:method-block}

\subsubsection{Noise Layers}

\label{sec:method-noise}
To enhance robustness of network against various types of distortions, we design a differentiable noise layer, which is inserted between the encoder and decoder during  training process of the network. The differentiable noise layer utilized during training consists of three types of noise sources:
\begin{equation}
    N_{pool} = \left\{\textit{Identity, Speckle, Gaussian, Resample}\right\}.
    \label{eq:train_pool}
\end{equation}
During training, for each batch of audio, we randomly select one of these noise types from $N_{pool}$ to apply as perturbation.

In order to assess the model robustness to different types of noise, we utilize a noise layer during testing that includes
\begin{align}
\notag
    N^{test}_{pool} =& \{\textit{Identity, Speckle, Gaussian, Resample,} \\ & \textit{MP3~compression, 8-bit~reduction}\}.
    \label{eq:test_pool}
\end{align}
We provide two versions of the training process: one with the noise layer incorporated and one without. In the training of the model with the noise layer, we build upon a pre-trained model without the noise layer. We decouple the forward and backward processes of the pre-trained model, fixing the parameters of the forward concealment process, while only training the parameters of the backward reveal process. This approach helps ensure that the concealed information remains imperceptible and allows the reveal network to adapt to the presence of noise.

\subsection{Security Mechanism}
\label{sec:method-security}

To ensure the security and integrity of the hidden audio data, our method integrates a scrambling encryption mechanism into the data hiding process. To achieve this, we segment the original secret audio \(x^{orig}_{secret}\) into very short fragments and then scramble all segments using a pseudo-random number generator (PRNG) \(\pi\) as an encryption measure. The seed \(\sigma\) used for the PRNG is securely shared between the sender and receiver to ensure that only authorized parties can reconstruct the original secret audio.

\subsubsection{Encryption Function}
The encryption of \(x^{orig}_{secret}\) occurs before hiding. We first divide \(x^{orig}_{secret}\) into \(n\) segments:
\begin{equation}
    x^{orig}_{secret} = \left[x_1, x_2, \dots, x_n\right].
\end{equation}
Then, the encrypted secret audio \(x_{secret}\) can be represented as:
\begin{equation}
    x_{secret} = \left[x_{\pi(\sigma)(1)}, x_{\pi(\sigma)(2)}, \dots, x_{\pi(\sigma)(n)}\right],
\end{equation}
where \(\pi(\sigma)(i)\) indicates the new position of the \(i\)-th element of the original array after scrambling.
This scrambling acts as an encryption step, effectively obfuscating the audio segments before they are embedded into the cover audio.

\subsubsection{Decryption Function}
The decryption process is the reverse of the encryption process. First, the encrypted secret audio \(x^{rev}_{secret}\) is revealed from the stego audio and divided into \(n\) segments:
\begin{equation}
    x^{rev}_{secret} = \left[x'_1, x'_2, \dots, x'_n\right].
\end{equation}
Subsequently, the original order and content of the secret audio are restored using the inverse of the pseudo-random scrambling sequence used for encryption:
\begin{equation}
    x^{restored}_{secret} = \left[x'_{\pi^{-1}(\sigma)(1)}, x'_{\pi^{-1}(\sigma)(2)}, \dots, x'_{\pi^{-1}(\sigma)(n)}\right],
\end{equation}
where \(\pi^{-1}(\sigma)(i)\) is the inverse mapping of the scrambling, indicating the original position that the \(i\)-th element should be restored to after scrambling.

\subsubsection{Security Analysis}
To quantify the security of our method, we calculate the probability \(P\) that an attacker successfully extracts the secret message. The security of the encryption depends on the unpredictability of the pseudo-random sequence, assuming only the sender and intended receiver know this sequence. Thus, the probability \(P\) that an attacker successfully extracts the secret message can be calculated using the number of possible permutations of \(n\) segments:
\begin{equation}
    P = \frac{1}{n!}.
\end{equation}
For an audio segment lasting 2 seconds, with \(n=100\) resulting in segment lengths of 20ms, the probability of the secret being unauthorizedly extracted is \(1.07 \times 10^{-158}\). 20ms is less than the duration of a phoneme in English, and semantically meaningful words usually consist of one or more phonemes; therefore, such short audio segments cannot convey any semantic information. The short duration of audio segments and the low decryption probability ensure that the secret message will not be extracted by unauthorized parties. 

Building on this foundation, our security mechanism not only ensures the confidentiality of the hidden data but also its integrity, as any unauthorized attempt to modify or reorder the data without the correct sequence would result in a noticeable degradation of the recovered audio quality, alerting the receiver to potential tampering. Additionally, the encryption mechanism is designed to be plug-and-play, allowing it to be seamlessly integrated with different segmentations $n$ and keys $\sigma$. This flexibility ensures that no additional training or complex configuration adjustments are necessary when changing the number of segments or updating the encryption key.

\subsection{Loss Function}
\label{sec:method-loss}

The overall loss function comprises two components: one part is the guidance loss, intended to encourage the stego audio to resemble the original cover audio and the distribution of the redundant information $r$ to resemble an isotropic spherical Gaussian distribution. Another part is the reconstruction loss, designed to encourage the recovered secret audio and cover audio to resemble the original secret audio and cover audio.

\subsubsection{Concealing Loss}
The forward concealment process begins with encrypting the original secret audio \(x^{orig}_{secret}\) to produce \(x_{secret}\).  Subsequently, we generate the stego audio $x_{stego}$ and the redundant information $r$ from the cover audio $x_{cover}$ and the encrypted secret audio $x_{secret}$. For $x_{stego}$, we encourage it to be similar to $x_{cover}$, resulting in the following expression:
\begin{equation}
    \mathcal{L}_{stego}\left(\theta\right) = \sum^{N}_{n = 1} \ell_{C}\left(x^{\left(n\right)}_{cover}, x^{\left(n\right)}_{stego}\right),
\end{equation}
where $\theta$ represents the network parameters, and $x_{stego}$ is the output of the function $f_\theta(x_{cover}, x_{secret})$. The function $\ell_{C}$ is used to measure the similarity between $x_{cover}$ and $x_{stego}$.

For the redundant information $r$, we aim to make $r$ as close as possible to a specified distribution $p$, which is independent of the case. This facilitates the recovery stage, where the auxiliary variable $z$, with a distribution similar to that of the redundant information $r$, is used to recover $x_{secret}$ in the absence of redundant information $r$. Therefore, we compute the log-likelihood of the redundant information $r$ conforming to the specified distribution $p$:
\begin{equation}
    \mathcal{L}_{z}\left(\theta\right) = -\sum^{N}_{n = 1} \log\left(p\left(r^{\left(n\right)}\right)\right).
\end{equation}

Without loss of generality, we employ a spherical multivariate Gaussian distribution $p(r) = N(r; 0, I)$ as the specified distribution.

The final loss in the concealment stage is a weighted sum of these two loss components:
\begin{equation}
    \mathcal{L}_{conceal}\left(\theta\right) = \lambda_{stego} \mathcal{L}_{stego}\left(\theta\right) + \lambda_{z} \mathcal{L}_{z}\left(\theta\right),
\end{equation}
where $\lambda_{stego}$ and $\lambda_{z}$ are weights used to balance the different loss terms.

\subsubsection{Revealing Loss}

The backward recovery process involves the recovery of the cover audio $x^{rev}_{cover}$ and encrypted secret $x^{rev}_{secret}$ from the stego audio $x_{stego}$ and the auxiliary variable $z$. Following this, we decrypt $x^{rev}_{secret}$ to obtain the secret audio $x^{restored}_{secret}$. We encourage $x^{rev}_{cover}$ to be similar to $x_{cover}$ and the decrypted secret $x^{restored}_{secret}$ to be similar to $x^{orig}_{secret}$, leading to the following expressions:
\begin{equation}
    \mathcal{L}_{cover}\left(\theta\right) = \sum ^{N}_{n = 1} \ell_{R}\left(x^{\left(n\right)}_{cover}, x^{rev\left(n\right)}_{cover}\right),
\end{equation}
\begin{equation}
    \mathcal{L}_{secret}\left(\theta\right) = \sum ^{N}_{n = 1} \ell_{R}\left(x^{orig\left(n\right)}_{secret}, x^{restored\left(n\right)}_{secret}\right),
\end{equation}
where the function \(l_R(x_1, x_2)\) is employed to gauge the similarity between \(x_1\) and \(x_2\).
The final revealing loss is a weighted sum of these two loss components:
\begin{equation}
    \mathcal{L}_{reveal}\left(\theta\right) = \lambda_{cover} \mathcal{L}_{cover}\left(\theta\right) + \lambda_{secret} \mathcal{L}_{secret}\left(\theta\right),
\end{equation}
where $\lambda_{cover}$ and $\lambda_{secret}$ are weights used to balance the different loss terms.

\subsubsection{Time-frequency-domain Loss}
In both the forward and backward processes, we employ $\mathcal{L}_{stego}$, $\mathcal{L}_{cover}$, and $\mathcal{L}_{secret}$ to constrain the stego audio, recovered cover, and recovered secret. To ensure that the stego audio closely matches the cover audio in terms of auditory perception, we utilize a multi-resolution STFT loss, which is originally used in audio generation tasks~\cite{yamamoto2020parallel}. This loss helps in generating natural-sounding audio. We denote the set of STFT window sizes used for spectral computation as $W = [w_1, w_2, \ldots, w_k]$, where $k$ is the number of window sizes in this set, and $w$ represents the window size used in STFT calculations. Given an STFT window size of $m$, the amplitude spectrum of the audio is represented as $S^{m}(\cdot)$. The multi-resolution STFT loss is defined as:
\begin{equation}
    \mathcal{L}_{STFT}\left(x_1, x_2\right) = \sum_{w \in W} \left\|S^{w}\left(x_1\right) - S^{w}\left(x_2\right)\right\|.
\end{equation}

To ensure that the outputs of the concealment and reveal networks align with human auditory perception, we apply 
 the $\mathcal{L}_{STFT}$ loss function for $\mathcal{L}_{stego}$, $\mathcal{L}_{cover}$, and $\mathcal{L}_{secret}$. During the initial pre-training phase, we use a single-resolution STFT loss with a window size set to $W = [1024]$. After the network has reached convergence, we adjust the window sizes to $W = [256, 512, 1024, 2048]$, enhancing the naturalness of the produced audio.

\subsubsection{Total Loss Function}
The overall loss function $\mathcal{L}_{total}(\theta)$ is the sum of two components, $\mathcal{L}_{conceal}(\theta)$ and $\mathcal{L}_{reveal}(\theta)$, and the final total loss function is as follows:
\begin{align}
\notag
    \mathcal{L}_{total}\left(\theta\right) = &\lambda_{stego} \mathcal{L}_{stego}\left(\theta\right) + \lambda_{z} \mathcal{L}_{z}\left(\theta\right) + \\ &\lambda_{cover} \mathcal{L}_{cover}\left(\theta\right) + \lambda_{secret} \mathcal{L}_{secret}\left(\theta\right),
\end{align}
where $\lambda_{stego}$, $\lambda_{z}$, $\lambda_{cover}$, and $\lambda_{secret}$ are weights used to balance the contributions of the different loss components.

\section{Experiments}
\label{sec:experiment}
In this section, we initially present the datasets used and our experimental setup. Then, we evaluate the performance of our method from both objective and subjective perspectives, comparing it with existing state-of-the-art methods. Objective experiments primarily encompass fidelity, robustness, and generalization aspects, while subjective experiments involve imperceptibility tests. Finally, we investigate the impact of individual components constituting our method.

\subsection{Experimental Setup}
\begin{table*}[htbp]
  \centering
  \caption{Benchmark Comparisons on Different Datasets, with the Best Results in Bold and the Second Bests Underlined}
    \begin{tabular}{c|cccc|cccc|cccc}
    \toprule
    \multicolumn{13}{c}{Cover/Stego Speech Pair} \\
    \midrule
    \multirow{2}[4]{*}{Method} & \multicolumn{4}{c|}{LibriSpeech-16kHz} & \multicolumn{4}{c|}{VCTK-16kHz} & \multicolumn{4}{c}{VCTK-24kHz} \\
    \cmidrule{2-13}      & SNR(dB)↑ & LSD↓  & PESQ↑ & SECS↑  & SNR(dB)↑ & LSD↓  & PESQ↑ & SECS↑  & SNR(dB)↑ & LSD↓  & PESQ↑ & SECS↑ \\
    \midrule
    LSB   & 3.59  & 1.88  & 1.12  & 0.57  & 3.44  & 1.85  & 1.11  & 0.52  & 3.45  & 1.96  & 1.08  & 0.53  \\
    Freq. Chop & 1.97  & 1.79  & 1.12  & 0.80  & 2.23  & 1.78  & 1.13  & 0.80  & 2.28  & 1.91  & 1.09  & 0.84  \\
    Hide\&Speak & \underline{26.88}  & \underline{0.79}  & \underline{3.28}  & \textbf{0.96} & \underline{27.28}  & \underline{0.74}  & \underline{3.14}  & \textbf{0.96} & \underline{27.51}  & \underline{0.78}  & \underline{3.02}  & \underline{0.96}  \\
    \METHODNAME~  & \textbf{27.03} & \textbf{0.62} & \textbf{3.89} & \underline{0.95}  & \textbf{27.52} & \textbf{0.61} & \textbf{4.00} & \textbf{0.96} & \textbf{27.52} & \textbf{0.65} & \textbf{3.65} & \textbf{0.97} \\
    \bottomrule
    \end{tabular}%

  \vspace{4pt} 
  \centering

    \begin{tabular}{c|cccc|cccc|cccc}
    \toprule
    \multicolumn{13}{c}{Secret/Recovery Speech Pair} \\
    \midrule
    \multirow{2}[4]{*}{Method} & \multicolumn{4}{c|}{LibriSpeech-16kHz} & \multicolumn{4}{c|}{VCTK-16kHz} & \multicolumn{4}{c}{VCTK-24kHz} \\
    \cmidrule{2-13}      & SNR(dB)↑ & LSD↓  & PESQ↑ & SECS↑  & SNR(dB)↑ & LSD↓  & PESQ↑ & SECS↑  & SNR(dB)↑ & LSD↓  & PESQ↑ & SECS↑ \\
    \midrule
    LSB   & -10.31  & 3.36  & 1.03  & \underline{0.41}  & -9.64  & 3.29  & 1.03  & 0.37  & -9.60  & 3.53  & 1.03  & 0.40  \\
    Freq. Chop & \underline{-1.93}  & 1.82  & \textbf{1.67}  & 0.22  & \underline{-1.89}  & 1.77  & \underline{1.58}  & 0.17  & \underline{-1.82}  & 1.73  & \underline{1.48}  & 0.29  \\
    Hide\&Speak & -2.60  & \underline{1.47}  & 1.24  & \underline{0.41}  & -2.65  & \underline{1.40}  & 1.22  & \underline{0.40}  & -2.68  & \underline{1.44}  & 1.21  & \underline{0.81}  \\
    \METHODNAME~  & \textbf{11.16} & \textbf{1.32} & \underline{1.49} & \textbf{0.69} & \textbf{12.86} & \textbf{1.21} & \textbf{1.61} & \textbf{0.79} & \textbf{16.79} & \textbf{1.05} & \textbf{1.70} & \textbf{0.91} \\
    \bottomrule
    \end{tabular}%

  \label{tab:fidelity}%
\end{table*}%
\subsubsection{Datasets and Settings}
We employ the standard training/validation/testing split and conduct training and testing on well-known speech datasets, specifically the VCTK dataset and the LibriSpeech dataset. The VCTK dataset includes speech data from 110 English speakers with different accents, each reading approximately 400 sentences. To evaluate the generalization of different methodologies across unseen speakers, we earmark speech data from four individuals—specifically, speakers 247, 305, 307, and 374—as the test subset to represent unseen speakers. On the other hand, LibriSpeech consists of 1000 hours of read English speech. All audio excerpts are uniformly resampled to both 16 kHz and 24 kHz to ensure consistency in our evaluations.

Our training examples are generated by randomly selecting one utterance as the cover and another utterance as the secret. Consequently, the pairing of covers and secrets is not fixed and could involve different speakers. To ensure consistency and uniformity in data handling, all audio is segmented into 2-second clips. In the scrambling encryption process, the parameter $n$ is set to 100, allowing each 2-second audio segment to be divided into 100 parts of 20 ms each. For samples with a 16 kHz sampling rate, the batch size is set to 4, while for samples with a 24 kHz sampling rate, the batch size is set to 2. The number of invertible blocks ($M$) used in the INN is set to 16. The parameters $\lambda_{stego}$, $\lambda_{z}$, $\lambda_{cover}$, and $\lambda_{secret}$ are set to 2.0, 10.0, 1.0, and 1.0, respectively. All models are trained using the Adam optimizer for 80 epochs, with an initial learning rate of $10^{-4.5}$ and a decay factor of 10 every 20 epochs.

\subsubsection{Benchmarks}
To validate the effectiveness of our approach, we compared our method against state-of-the-art (SOTA) audio steganography techniques, which include a traditional 8-bit LSB method, a naïve baseline approach called Frequency Chop, and a deep-learning-based method: Hide \& Speak~\cite{kreuk2019hide}. In Frequency Chop, we concatenated the upper frequency portion of $x_{secret}$ with the lower frequency portion of $x_{cover}$ to create $x_{stego}$. For a fair comparison, we trained our model and the Hide 
\& Speak~\cite{kreuk2019hide} model using the same dataset.

\subsubsection{Metrics}
To assess the performance of our method in terms of fidelity and robustness, we utilized both objective and subjective metrics to assess the quality of the cover/stego pairs and secret/recovery pairs.  The objective metrics employed include Signal-to-noise ratio (SNR), Log-Spectal Distortion(LSD), Perceptual Evaluation of Speech Quality (PESQ), and Speaker Embedding Cosine Similarity (SECS). Their definitions are as follows:
\begin{itemize}
\item SNR: For objective evaluation, we assessed the fidelity of both the stego audio and recovered secret audio using the Signal-to-noise ratio (SNR).  For audio samples $X$ and $Y$, each comprising $n$ samples, the SNR is defined as follows:
\begin{equation}
\textrm{SNR} = 10 \cdot \log_{10}\left(\frac{\sum_{i=1}^n X_i^2}{\sum_{i=1}^n \left(X_i - Y_i\right)^2}\right),
\end{equation}
where $X_i$ and $Y_i$ represent the audio signal values at the i-th sample point.

\item LSD: Given that SNR is calculated in the time domain, and frequency-domain metrics are often more closely related to human auditory perception, we also used the frequency-domain metric known as Log-Spectral Distance (LSD) to measure the fidelity of our method. This metric calculates the distance between the log-spectral of the stego audio and the cover audio. The log-spectral distance is defined as follows: 
\begin{equation}
\textrm{LSD} = \left\{\frac{1}{N} \sum_{n=1}^N \left[\log P(X)_n - \log P(Y)_n\right]^p\right\}^{1/p},
\end{equation}
where $P\left(X\right)$ and ${P}\left(Y\right)$ are power spectra in discrete space.
\begin{figure*}
    \centering
    \includegraphics[width=0.9\linewidth]{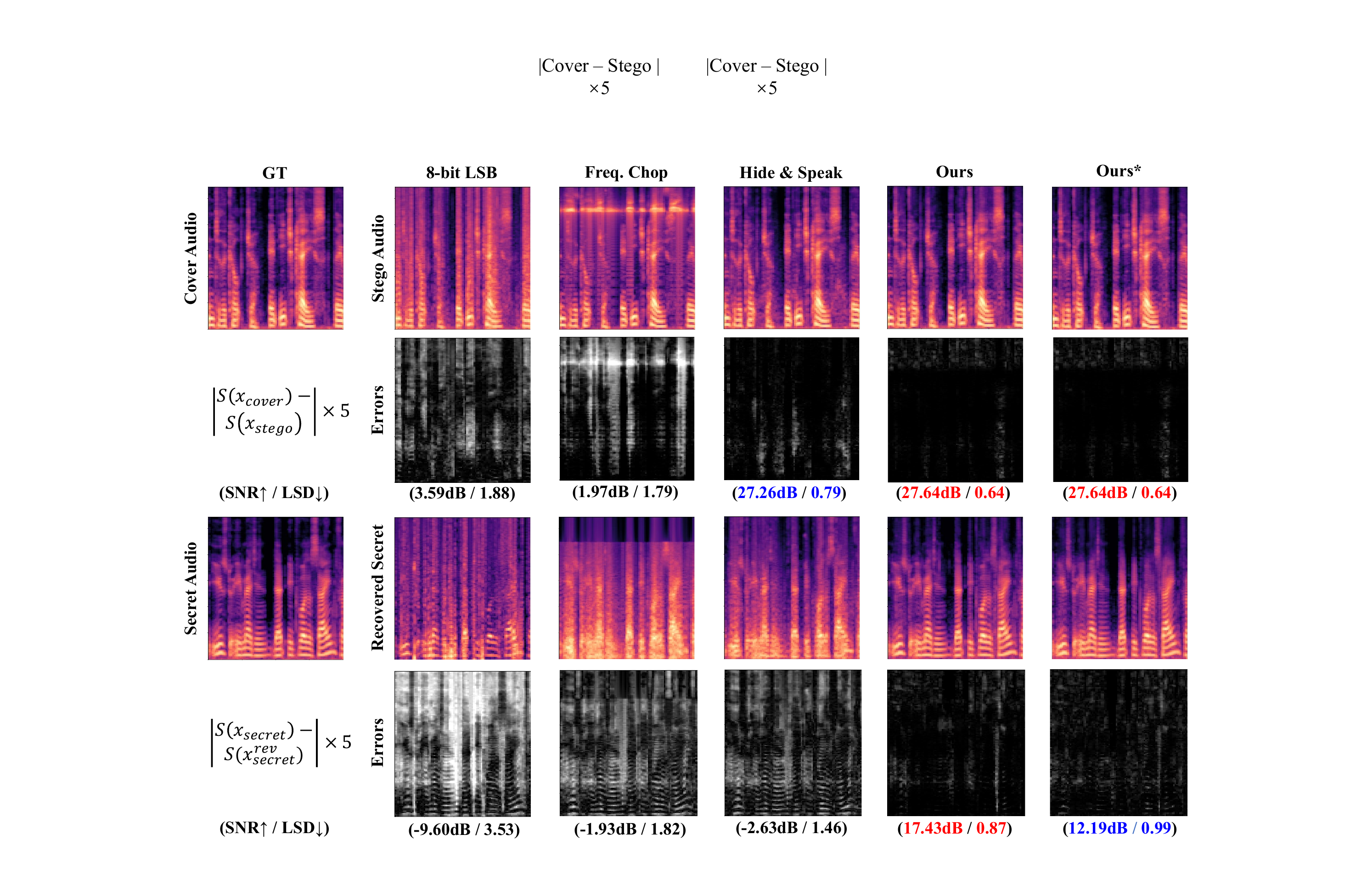}
    \caption{Visualization of Mel-spectrograms. The hiding operations of LSB, Frequency Chop, Hide\&Speak produce visible artifacts on recovered secret audio, whereas our method shows no such artifacts. The asterisk (*)  denotes models with noise layer added during training process. Best-performing results are highlighted in red, while second-best results are indicated in blue.}
    \label{fig:visual}
\end{figure*}

\item PESQ: Perceptual Evaluation of Speech Quality (PESQ) is a widely used standard for assessing the quality of processed speech signals in speech processing and audio evaluation. It measures the perceptual similarity between the original and degraded speech, offering a quantitative score to indicate the level of degradation from audio processing. The PESQ score ranges from -0.5 to 4.5, where higher scores denote greater fidelity to the original speech and better perceived quality. 
\item SECS: Speaker Embedding Cosine Similarity (SECS) measures the degree of similarity between the speaker embeddings extracted from the audio samples. A significant advantage of using speech as secret information compared to text is its inclusion of speaker identity information. To evaluate how well the recovered secret retains this speaker identity, we used cosine-similarity scores with speaker embeddings. We utilize a pre-trained speaker verification model called ECAPA-TDNN~\cite{desplanques2020ecapa} to extract speaker embeddings and calculate cosine similarity scores between embedding pairs. ECAPA-TDNN, based on a time delay neural network (TDNN), is a state-of-the-art voiceprint extractor. It achieves 0.17\% Equal Error Rate (EER) on the VCTK dataset and 0.19\% EER on the LibriSpeech dataset, indicating its strong ability to distinguish between different speakers. The SECS calculation between audio samples \(X\) and \(Y\) is determined by the cosine similarity formula:
\begin{equation}
\textrm{SECS}\left(X, Y\right) = \frac{{V\left(X\right) \cdot V\left(Y\right)}}{{\|V\left(X\right)\| \|V\left(Y\right)\|}},
\end{equation}
where \(V(\cdot)\) signifies the embedding function responsible for extracting speaker embeddings.
\end{itemize}

Higher values of SNR, PESQ, SECS and lower values of LSD suggest superior audio quality. In terms of subjective assessment, we employed ABX test.  If the test outcome is close to 50\%, this implies that the stego audio and the cover audio are indistinguishable to listeners, thus signifying the imperceptibility of the hidden information.

\subsection{Objective Evaluation}
\subsubsection{Fidelity}

Table~\ref{tab:fidelity} provides a comparison of the fidelity for our method against other approaches using SNR, LSD, PESQ, and SECS metrics on the Librispeech and VCTK datasets. The data clearly demonstrate that our method generally outperforms others in most cases across all metrics for the cover/stego audio pairs, and significantly excels in the secret/recovery audio pairs. 

With respect to SNR, our cover/stego audio pairs show enhancements of 0.15 dB, 0.01 dB, and 0.01 dB over the second-best results on the LibriSpeech, VCTK-16kHz, and VCTK-24kHz datasets, respectively. Meanwhile, our secret/recovery audio pairs demonstrate substantial improvements of 13.76 dB, 15.51 dB, and 19.47 dB, respectively. It is evident that the traditional LSB method exhibits subpar hiding quality due to its fixed and limited capacity. For example, the LSB and Frequency Chop methods manage SNR values of only 3.59 dB and 1.97 dB for the cover/stego audio pairs, respectively, which are over 20 dB lower than those achieved by our method. Such low SNR levels are easily detectable by the human ear, marking a clear failure in effective audio concealment. Compared to other deep-learning-based methods, our approach achieves markedly superior quality in the recovered secret audio.

The advantages of our method are further evidenced in additional metrics such as LSD, PESQ and SECS. Notably, a higher SECS indicates that the recovered secret audio retains more speaker identity information, showcasing the effectiveness of our method in preserving speaker characteristics within the secret audio. Additionally, our time-domain approach not only outperforms in time-domain metrics like SNR but also demonstrates better performance in frequency-domain metrics like LSD compared to previous frequency-domain methods. This indicates that the time-frequency loss integrated into our model promotes an understanding of the interrelations between the frequency and time domains, enhancing overall model performance.

To further compare speech quality, we visualize the obtained stego audio and recovered secret audio from different methods, as shown in Figure~\ref{fig:visual}. It is evident that both our method and Hide \& Speak~\cite{kreuk2019hide} produce  stego audio that is indistinguishable from the cover audio. However, in terms of the recovered secret audio, our method exhibits cleaner high-frequency components, while the high-frequency portions of recovered secret audio obtained from~\cite{kreuk2019hide} display noticeable artifacts. This once again demonstrates the superior fidelity achieved by our method.\footnote{Our demo is available at \url{https://cyberrrange.github.io/project/wavinwav}.}

\subsubsection{Robustness}

\renewcommand{\arraystretch}{1.15}
\begin{table}[htbp]
    \centering
    \caption{Comparison of Results Obtained by Applying Different Noises to Stego Audio in the VCTK Dataset with Benchmark}
    \setlength{\tabcolsep}{1.75mm}{\begin{tabular}{ccccccc}
    \toprule
    \multicolumn{2}{c}{\multirow{2}{*}{}}            & \multirow{2}{*}{Method} & \multicolumn{4}{c}{Metrics}                                    \\ \cline{4-7} 
    \multicolumn{2}{c}{}                             &                         & SNR(dB)↑       & LSD↓          & PESQ↑         & SECS↑         \\ \hline
    \multicolumn{2}{c}{\multirow{3}{*}{Cover/Stego}} & Hide\&Speak             & 27.51          & 0.78          & 3.02          & 0.96          \\
    \multicolumn{2}{c}{}                             & \METHODNAME~                    & \textbf{27.52} & \textbf{0.65} & \textbf{3.65} & \textbf{0.97} \\
    \multicolumn{2}{c}{}                             & \METHODNAME~*                   & \textbf{27.52} & \textbf{0.65} & \textbf{3.65} & \textbf{0.97} \\ \hline
    \multicolumn{1}{c}{\multirow{27}{*}{\rotatebox{90}{Secret/Recovery}}} &
      \multirow{3}{*}{Identity} &
      Hide\&Speak &
      -2.68 &
      1.44 &
      1.21 &
      0.39 \\
    \multicolumn{1}{c}{} &                          & \METHODNAME~                    & \textbf{16.79} & \textbf{1.05} & \textbf{1.70} & \textbf{0.91} \\
    \multicolumn{1}{c}{} &                          & \METHODNAME~*                   & 11.14          & 1.12          & 1.48          & 0.76          \\ \cline{3-7} 
    \multicolumn{1}{c}{} &
      \multirow{3}{*}{\begin{tabular}[c]{@{}c@{}}Down-\\      Sampling\end{tabular}} &
      Hide\&Speak &
      -2.79 &
      1.72 &
      1.10 &
      0.35 \\
    \multicolumn{1}{c}{} &                          & \METHODNAME~                    & 4.42           & 1.56          & 1.05          & 0.46          \\
    \multicolumn{1}{c}{} &                          & \METHODNAME~*                   & \textbf{7.67}  & \textbf{1.25} & \textbf{1.21} & \textbf{0.49} \\ \cline{3-7} 
    \multicolumn{1}{c}{} &
      \multirow{3}{*}{Gaussian} &
      Hide\&Speak &
      -2.81 &
      2.07 &
      \textbf{1.07} &
      0.26 \\
    \multicolumn{1}{c}{} &                          & \METHODNAME~                    & -7.53          & 2.47          & 1.04          & 0.25          \\
    \multicolumn{1}{c}{} &                          & \METHODNAME~*                   & \textbf{2.12}  & \textbf{1.31} & 1.06          & \textbf{0.27} \\ \cline{3-7} 
    \multicolumn{1}{c}{} & \multirow{3}{*}{Speckle} & Hide\&Speak             & -2.72          & 1.65          & 1.14          & 0.36          \\
    \multicolumn{1}{c}{} &                          & \METHODNAME~                    & 2.27           & 1.68          & 1.08          & \textbf{0.67} \\
    \multicolumn{1}{c}{} &                          & \METHODNAME~*                   & \textbf{6.79}  & \textbf{1.20} & \textbf{1.21} & 0.58          \\ \cline{3-7} 
    \multicolumn{1}{c}{} &
      \multirow{3}{*}{\begin{tabular}[c]{@{}c@{}}MP3\\      128kbps\end{tabular}} &
      Hide\&Speak &
      -2.50 &
      1.52 &
      1.16 &
      0.38 \\
    \multicolumn{1}{c}{} &                          & \METHODNAME~                    & \textbf{9.48}  & 1.40          & 1.18          & \textbf{0.77} \\
    \multicolumn{1}{c}{} &                          & \METHODNAME~*                   & 8.91           & \textbf{1.25} & \textbf{1.27} & 0.66          \\ \cline{3-7} 
    \multicolumn{1}{c}{} &
      \multirow{3}{*}{\begin{tabular}[c]{@{}c@{}}MP3\\      96kbps\end{tabular}} &
      Hide\&Speak &
      -2.58 &
      1.67 &
      1.12 &
      0.35 \\
    \multicolumn{1}{c}{} &                          & \METHODNAME~                    & 4.36           & 1.57          & 1.09          & \textbf{0.66} \\
    \multicolumn{1}{c}{} &                          & \METHODNAME~*                   & \textbf{6.59}  & \textbf{1.30} & \textbf{1.17} & 0.57          \\ \cline{3-7} 
    \multicolumn{1}{c}{} &
      \multirow{3}{*}{\begin{tabular}[c]{@{}c@{}}MP3\\      64kbps\end{tabular}} &
      Hide\&Speak &
      -3.27 &
      2.08 &
      \textbf{1.07} &
      0.27 \\
    \multicolumn{1}{c}{} &                          & \METHODNAME~                    & -1.37          & 1.85          & 1.05          & \textbf{0.40} \\
    \multicolumn{1}{c}{} &                          & \METHODNAME~*                   & \textbf{2.39}  & \textbf{1.44} & \textbf{1.07} & 0.39          \\ \cline{3-7} 
    \multicolumn{1}{c}{} &
      \multirow{3}{*}{\begin{tabular}[c]{@{}c@{}}MP3\\      32kbps\end{tabular}} &
      Hide\&Speak &
      -4.49 &
      2.23 &
      \textbf{1.05} &
      \textbf{0.14} \\
    \multicolumn{1}{c}{} &                          & \METHODNAME~                    & -3.91          & 1.84          & 1.04          & 0.00          \\
    \multicolumn{1}{c}{} &                          & \METHODNAME~*                   & \textbf{-3.16} & \textbf{1.68} & 1.04          & -0.02         \\ \cline{3-7} 
    \multicolumn{1}{c}{} &
      \multirow{3}{*}{\begin{tabular}[c]{@{}c@{}}8-bit\\      Reduction\end{tabular}} &
      Hide\&Speak &
      -2.87 &
      2.10 &
      1.05 &
      0.23 \\
    \multicolumn{1}{c}{} &                          & \METHODNAME~                    & -6.03          & 2.32          & 1.04          & 0.23          \\
    \multicolumn{1}{c}{} &                          & \METHODNAME~*                   & \textbf{1.62}  & \textbf{1.35} & \textbf{1.06} & \textbf{0.26} \\ 
    \bottomrule
    \multicolumn{7}{l}{\thead{* The asterisk denotes models with noise layer added during the training\\
    process. The best results are indicated in bold.~~~~~~~~~~~~~~~~~~~~~~~~~~~}}
    \end{tabular}}
  \label{tab:robustness}%
\end{table}
\renewcommand{\arraystretch}{1}

In addition to fidelity under noise-free conditions, the performance under noisy conditions is also crucial. We applied different channel distortions and compression techniques to the stego audio, and Table~\ref{tab:robustness} compares the performance of different methods under these distortion conditions, using the noise pool consistent with Equation~\ref{eq:test_pool}. Gaussian and Speckle noises were applied with $\sigma = 0.001$ and $\sigma = 0.01$, respectively. DownSampling distortion resampled the 24kHz stego audio to 16kHz. MP3 compression was applied at rates of 128kbps, 96kbps, 64kbps, and 32kbps. For our method, we employed two settings: one with and one without the use of the noise pool consistent with Equation~\ref{eq:train_pool} during training, denoted as \METHODNAME~* and \METHODNAME~ respectively. Since the 8-bit LSB and Frequency Chop baselines have already demonstrated their inability to effectively conceal audio in noiseless conditions, for the sake of brevity, we omit further comparisons with these two baselines at this stage.

From Table~\ref{tab:robustness}, it is evident that in most cases, our method outperforms others. Under weaker noise intensity and noise-free conditions, \METHODNAME~ shows better performance, whereas under higher noise intensity, \METHODNAME~* performs better, indicating a trade-off between robustness and transparency. Notably, \METHODNAME~ achieved some robustness against MP3 compression and DownSampling noise even without using the noise layer during training. This is attributed to our use of time-frequency loss during training, enabling the model to learn information hiding and extraction in the frequency domain, 
while DownSampling and MP3 compression primarily preserve the frequency-domain characteristics of the audio. Additionally, the auxiliary variable $z$ is randomly sampled from a gaussian distribution, contributing to the  model robustness.

For MP3 compression below 64kbps, Gaussian noise, and 8-bit precision reduction, our method exhibited relatively weaker robustness. However, it still outperformed other methods in most metrics. The observed outcomes stem from the intricate balance between robustness and transparency within our approach. Our method exhibits minimal alterations in both the frequency and time domains of the cover audio, ensuring a balance between concealing the information and maintaining its fidelity, while the introduction of Gaussian noise and a reduction in precision by 8 bits exert a pronounced influence on the time-domain characteristics of the stego audio. Furthermore, as the bit rate diminishes during MP3 compression, it eliminates a considerable amount of perceptual redundancy, consequently inflicting substantial damage on the imperceptible concealed information.

\subsubsection{Generalization Ability}

\begin{table*}[htbp]
  \centering
  \caption{Comparison of Results Obtained by Applying Different Methods to Unseen Speakers and Unseen Datasets}
    \begin{tabular}{c|ccc|ccc|ccc|ccc}
    \toprule
    \multirow{3}[6]{*}{Method} & \multicolumn{6}{c|}{Cover/Stego Speech Pair}  & \multicolumn{6}{c}{Secret/Recovery Speech Pair} \\
\cmidrule{2-13}          & \multicolumn{3}{c|}{Unseen Speakers (VCTK)} & \multicolumn{3}{c|}{Unseen Dataset (VCTK)} & \multicolumn{3}{c|}{Unseen Speakers (VCTK)} & \multicolumn{3}{c}{Unseen Dataset (VCTK)} \\
\cmidrule{2-13}          & SNR(dB)↑ & LSD↓  & SECS↑  & SNR(dB)↑ & LSD↓  & SECS↑  & SNR(dB)↑ & LSD↓  & SECS↑  & SNR(dB)↑ & LSD↓  & SECS↑ \\
    \midrule
    Hide\&Speak & \textbf{26.65} & 0.79  & \textbf{0.95} & 26.91 & 0.79  & \textbf{0.96} & -2.67  & 1.42  & 0.45  & -2.60  & 1.47  & 0.40  \\
    \METHODNAME~  & 26.25  & \textbf{0.62} & \textbf{0.95} & \textbf{27.83}  & \textbf{0.59} & \textbf{0.96}  & \textbf{11.11} & \textbf{1.10} & \textbf{0.78} & \textbf{12.60} & \textbf{1.23} & \textbf{0.76} \\
    \bottomrule
    \end{tabular}%
  \label{tab:general}%
\end{table*}%
The generalization performance of audio data hiding methods is crucial for their practical applicability. Our experiments primarily focus on assessing the generalization ability concerning unseen speakers and across different datasets. Specifically, to evaluate the  ability on unseen speakers of our model, we isolated four speakers from the VCTK dataset as the unseen speaker set while training on the remaining 106 speakers. For assessing generalization across datasets, we trained the model on the LibriSpeech dataset and tested it on the VCTK dataset. As depicted in Table~\ref{tab:general}, our approach achieves 26.25 SNR for cover/stego pairs and 11.11 SNR for secret/recovery pairs on unseen speakers. Additionally, on the unseen dataset, our method achieved 27.83 SNR for cover/stego pairs and 12.60 SNR for secret/recovery pairs, highlighting its robust generalization performance.
\begin{table*}[htbp]
  \centering
  \caption{Ablation Study on Time-domain Hiding, Time-frequency-domain Loss and Invertible Neural Network}
    \begin{tabular}{ccc|cccc|cccc}
    \toprule
    \multicolumn{1}{c}{\multirow{2}[4]{*}{INN}} & \multicolumn{1}{c}{\multirow{2}[4]{*}{\thead{Time-domain\\Hiding}}} & \multicolumn{1}{c|}{\multirow{2}[4]{*}{\thead{Time-frequency\\Loss}}} & \multicolumn{4}{c|}{Cover/Stego Speech Pair} & \multicolumn{4}{c}{Secret/Recovery Speech Pair} \\
    \cmidrule{4-11}\multicolumn{1}{c}{} &       &       & SNR(dB)↑ & LSD↓  & PESQ↑  & SECS↑ & SNR(dB)↑ & LSD↓  & PESQ↑  & SECS↑ \\
    \midrule
    \XSolidBrush     & \Checkmark     & \Checkmark     & 24.65  & 0.76  & 3.40  & 0.93  & 11.68  & 1.27  & 1.47  & 0.73  \\
    \Checkmark     & \XSolidBrush     & \Checkmark     & 13.57  & 0.92  & 1.54  & 0.71  & -2.49  & 1.28  & 1.11  & 0.23  \\
    \Checkmark     & \Checkmark     & \XSolidBrush     & 26.84  & 0.83  & 2.83  & 0.94  & 12.25  & 1.49  & 1.26  & 0.70  \\
    \Checkmark     & \Checkmark     & \Checkmark     & \textbf{27.52} & \textbf{0.61} & \textbf{4.00} & \textbf{0.96} & \textbf{12.86} & \textbf{1.21} & \textbf{1.61} & \textbf{0.79} \\
    \bottomrule
    \end{tabular}%

  \label{tab:ablation}%
\end{table*}%
\subsection{Subjective Detectability Test}
To verify that humans cannot detect the differences between cover and stego audio, we conducted ABX tests. For each test utterance, we provide two audio samples, A and B, to the volunteers participating in the test. One of these samples is the cover audio, and the other is the stego audio. Subsequently, we present an audio sample X, randomly chosen from A and B, and the participants have to decide whether X is the same as A or B. We generated 20 audio examples, and for each example, we recorded 20 responses, resulting in a total of 400 responses. Only 50.75\% of the stego audio could be distinguished from the cover by humans, which is close to random guessing (50\%). This suggests that the distortion caused by our method is nearly imperceptible to the human ear.
\comments{TODO: 是否补secret audio的主观MOS结果}

\subsection{Ablation Study}

\subsubsection{Effect of Time-domain Hiding}

To assess the impact of hiding in different domains, we maintained a comparable number of network parameters and substituted the network input and output with the amplitude spectrum of audio for comparison alongside our method. As shown in Table~\ref{tab:ablation}, hiding audio in the time domain resulted in a significant increase of 13.95 in SNR for cover/stego speech pairs and 15.35 in SNR for secret/recovery speech pairs. This effect stems from directly embedding and extracting in the time domain, which avoids distortions introduced to the stego audio spectrogram by the STFT-iSTFT transformation, thereby preventing the extraction of the secret spectrogram from a distorted stego spectrogram. Additionally, since the extracted secret audio is also in the time domain, there is no need to infer the phase spectrum from the amplitude spectrum, further circumventing severe phase distortions resulting from inferring the phase from distorted amplitude spectra.

\subsubsection{Effect of INN}
As an alternative to the INN architecture, we independently modeled the hiding and extraction processes using the traditional encoder-decoder framework. Specifically, we ensured a similar quantity of network parameters and constructed two U-Net architectures, akin to our experimental setup, as separate networks for the hiding and extraction processes. From Table~\ref{tab:ablation}, we observed that the INN-based network demonstrated superior overall performance. We attribute this to the intrinsic reversibility of the INN, which naturally fits the modeling of the inverse processes involved in information hiding and extraction. Hence, the utilization of INN yielded improved fidelity.

\subsubsection{Effect of Time-frequency-domain Loss Function}
In Table~\ref{tab:ablation}, we conducted a comparative analysis against \(\ell_2\) time-domain loss.  The outcomes revealed that the time-frequency loss contributed to an increase of 0.68 in SNR for cover/stego speech pairs and 0.61 in SNR for secret/recovery speech pairs. Moreover, there were more pronounced improvements in LSD and PESQ metrics, with a 1.17 enhancement in PESQ for cover/stego speech pairs and a 0.35 enhancement for secret/recovery speech pairs.  As LSD primarily reflects time-frequency characteristics and PESQ demonstrates a stronger alignment with time-frequency metrics than time-domain metrics, these outcomes align intuitively. However, the time-frequency loss also engendered enhancements in the time-domain metric SNR. This phenomenon might stem from the challenge posed by the time-domain loss,  making it harder for the model to capture the time-frequency correlations in the audio, thereby impeding convergence to the optimal solution.

\section{Conclusion }
\label{sec:conclusion}
In this paper, we introduce a novel approach named \METHODNAME~ for the end-to-end concealment and recovery of complete secret audio within cover audio. Our technique utilizes a novel flow-based invertible neural network that establishes a direct mapping between stego, cover, and secret audio. During network training, we introduce a time-frequency loss on time-domain signals to optimize the quality of both stego and secret audio. This strategy effectively reduces the loss of secret information typically caused by time-frequency transformations in previous methods, while retaining the benefits of time-frequency constraints to enhance the reversibility of message recovery. We also incorporate encryption technology into our system to protect the hidden data from unauthorized access. We conducted extensive qualitative and quantitative experiments on the VCTK and LibriSpeech datasets. Qualitative results show that modifications to the cover audio made by our method are imperceptible to human listeners. Quantitatively, both the stego audio and the recovered secret audio significantly outperform other state-of-the-art methods in various objective metrics, particularly with a notable 13-20dB improvement in SNR for the secret audio. Additionally, our method demonstrates robustness against multiple types of noise and performs well on unseen speakers and datasets. Ablation studies have confirmed the effectiveness of each component in our system.

However, there are still some limitations. Our method is vulnerable to strong noise attacks that can significantly affect minor modifications to the cover audio. Moreover, we have also not yet assessed how well our method resists steganalysis techniques. It is important to note that, like some recent work in this field~\cite{kreuk2019hide, xu2022robust, ke2024stegformer, geleta2022pixinwav, yang2019video, takahashi2021source}, our approach focuses on achieving perceptual transparency rather than resisting steganalysis, which has already become a common task. Future work could focus on enhancing system robustness against different kinds of noise and improving its resistance to steganalysis. Currently, most DNN-based audio hiding methods focus on the WAV format. Exploring the application of DNNs for hiding audio in other formats like MP3 remains an area ripe for investigation. Furthermore, given the impressive results of invertible neural networks in the image domain, further exploration of their application in audio hiding promises to be fruitful.




\bibliographystyle{IEEEtran}
\bibliography{ref}

\begin{thebibliography}{10}
\providecommand{\url}[1]{#1}
\csname url@samestyle\endcsname
\providecommand{\newblock}{\relax}
\providecommand{\bibinfo}[2]{#2}
\providecommand{\BIBentrySTDinterwordspacing}{\spaceskip=0pt\relax}
\providecommand{\BIBentryALTinterwordstretchfactor}{4}
\providecommand{\BIBentryALTinterwordspacing}{\spaceskip=\fontdimen2\font plus
\BIBentryALTinterwordstretchfactor\fontdimen3\font minus
  \fontdimen4\font\relax}
\providecommand{\BIBforeignlanguage}[2]{{%
\expandafter\ifx\csname l@#1\endcsname\relax
\typeout{** WARNING: IEEEtran.bst: No hyphenation pattern has been}%
\typeout{** loaded for the language `#1'. Using the pattern for}%
\typeout{** the default language instead.}%
\else
\language=\csname l@#1\endcsname
\fi
#2}}
\providecommand{\BIBdecl}{\relax}
\BIBdecl

\bibitem{chen2021distribution}
K.~Chen, H.~Zhou, H.~Zhao, D.~Chen, W.~Zhang, and N.~Yu,
  ``Distribution-preserving steganography based on text-to-speech generative
  models,'' \emph{IEEE Transactions on Dependable and Secure Computing},
  vol.~19, no.~5, pp. 3343--3356, 2021.

\bibitem{cuiicassp2020imginaudio}
W.~Cui, S.~Liu, F.~Jiang, Y.~Liu, and D.~Zhao, ``Multi-stage residual hiding
  for image-into-audio steganography,'' in \emph{ICASSP 2020 - 2020 IEEE
  International Conference on Acoustics, Speech and Signal Processing
  (ICASSP)}, 2020, pp. 2832--2836.

\bibitem{kreuk2019hide}
F.~Kreuk, Y.~Adi, B.~Raj, R.~Singh, and J.~Keshet, ``Hide and speak: Towards
  deep neural networks for speech steganography,'' \emph{arXiv preprint
  arXiv:1902.03083}, 2019.

\bibitem{baluja2017hiding}
S.~Baluja, ``Hiding images in plain sight: Deep steganography,'' \emph{Advances
  in neural information processing systems}, vol.~30, 2017.

\bibitem{zhu2018hidden}
J.~Zhu, R.~Kaplan, J.~Johnson, and L.~Fei-Fei, ``Hidden: Hiding data with deep
  networks,'' in \emph{Proceedings of the European conference on computer
  vision (ECCV)}, 2018, pp. 657--672.

\bibitem{yang2019video}
H.~Yang, H.~Ouyang, V.~Koltun, and Q.~Chen, ``Hiding video in audio via
  reversible generative models,'' in \emph{Proceedings of the IEEE/CVF
  International Conference on Computer Vision}, 2019, pp. 1100--1109.

\bibitem{yan2008reversible}
D.~Yan and R.~Wang, ``Reversible data hiding for audio based on prediction
  error expansion,'' in \emph{2008 International Conference on Intelligent
  Information Hiding and Multimedia Signal Processing}, 2008, pp. 249--252.

\bibitem{lsbascii}
A.~Mishra, P.~Johri, and A.~Mishra, ``Audio steganography using ascii code and
  ga,'' in \emph{2017 International Conference on Infocom Technologies and
  Unmanned Systems (Trends and Future Directions) (ICTUS)}, 2017, pp. 646--651.

\bibitem{lsbrsa}
A.~Gambhir and S.~Khara, ``Integrating rsa cryptography \& audio
  steganography,'' in \emph{2016 International Conference on Computing,
  Communication and Automation (ICCCA)}, 2016, pp. 481--484.

\bibitem{nassrullah2020enhancement}
H.~A. Nassrullah, W.~N. Flayyih, and M.~A. Nasrullah, ``Enhancement of lsb
  audio steganography based on carrier and message characteristics.'' \emph{J.
  Inf. Hiding Multim. Signal Process.}, vol.~11, no.~3, pp. 126--137, 2020.

\bibitem{fft2009fallahpour}
M.~Fallahpour and D.~Meg{\'i}as, ``High capacity method for real-time audio
  data hiding using the fft transform,'' 2009.

\bibitem{hongson2013dct}
C.~Hong-son, ``Research for a dct-based blind audio steganography algorithm,''
  \emph{Netinfo Security}, 2013.

\bibitem{dwt2002cvejic}
N.~Cvejic and T.~Seppanen, ``A wavelet domain lsb insertion algorithm for high
  capacity audio steganography,'' in \emph{Proceedings of 2002 IEEE 10th
  Digital Signal Processing Workshop, 2002 and the 2nd Signal Processing
  Education Workshop.}, 2002, pp. 53--55.

\bibitem{mp32016bazyar}
M.~Bazyar and R.~Sudirman, ``A new data embedding method for mpeg layer iii
  audio steganography,'' 2016.

\bibitem{ren2021silk}
Y.~Ren, S.~Zhong, W.~Tu, H.~Yang, and L.~Wang, ``A silk adaptive steganographic
  scheme based on minimizing distortion in pitch domain,'' \emph{IETE Technical
  Review}, vol.~38, no.~1, pp. 46--55, 2021.

\bibitem{ren2021aac}
Y.~Ren, S.~Cai, and L.~Wang, ``Secure aac steganography scheme based on
  multi-view statistical distortion (sofmvd),'' \emph{Journal of Information
  Security and Applications}, vol.~59, p. 102863, 2021.

\bibitem{tian2015voip}
H.~Tian, J.~Qin, Y.~Huang, Y.~Chen, T.~Wang, J.~Liu, and Y.~Cai, ``Optimal
  matrix embedding for voice-over-ip steganography,'' \emph{Signal Processing},
  vol. 117, pp. 33--43, 2015.

\bibitem{mazurczyk2008voip}
W.~Mazurczyk and J.~Lubacz, ``Lack—a voip steganographic method,''
  \emph{Telecommunication Systems}, vol.~45, pp. 153--163, 2008.

\bibitem{geleta2022pixinwav}
M.~Geleta, C.~Punti, K.~McGuinness, J.~Pons, C.~Canton, and X.~Giro-i Nieto,
  ``Pixinwav: Residual steganography for hiding pixels in audio,'' in
  \emph{ICASSP 2022-2022 IEEE International Conference on Acoustics, Speech and
  Signal Processing (ICASSP)}.\hskip 1em plus 0.5em minus 0.4em\relax IEEE,
  2022, pp. 2485--2489.

\bibitem{ros2023pixinwav2}
J.~Ros, M.~Geleta, J.~Pons, and X.~G. i~Nieto, ``Towards robust image-in-audio
  deep steganography,'' \emph{ArXiv}, vol. abs/2303.05007, 2023.

\bibitem{sridevi2009efficient}
R.~Sridevi, A.~Damodaram, and S.~Narasimham, ``Efficient method of audio
  steganography by modified lsb algorithm and strong encryption key with
  enhanced security.'' \emph{Journal of Theoretical \& Applied Information
  Technology}, vol.~5, no.~6, 2009.

\bibitem{chen2023imperceptible}
L.~Chen, R.~Wang, L.~Dong, and D.~Yan, ``Imperceptible adversarial audio
  steganography based on psychoacoustic model,'' \emph{Multimedia Tools and
  Applications}, vol.~82, no.~17, pp. 26\,451--26\,463, 2023.

\bibitem{djebbar2013phase}
F.~Djebbar, B.~Ayad, K.~Abed-Meraim, and H.~Hamam, ``Unified phase and
  magnitude speech spectra data hiding algorithm,'' \emph{Secur. Commun.
  Networks}, vol.~6, pp. 961--971, 2013.

\bibitem{echo2001}
H.~O. Oh, J.~W. Seok, J.~W. Hong, and D.~H. Youn, ``New echo embedding
  technique for robust and imperceptible audio watermarking,'' in \emph{2001
  IEEE International Conference on Acoustics, Speech, and Signal Processing.
  Proceedings (Cat. No.01CH37221)}, vol.~3, 2001, pp. 1341--1344 vol.3.

\bibitem{spread2006}
H.~Matsuoka, ``Spread spectrum audio steganography using sub-band phase
  shifting,'' in \emph{2006 International Conference on Intelligent Information
  Hiding and Multimedia}, 2006, pp. 3--6.

\bibitem{cvejic2004algorithms}
N.~Cvejic, ``Algorithms for audio watermarking and steganography,'' 2004.

\bibitem{prenger2019waveglow}
R.~Prenger, R.~Valle, and B.~Catanzaro, ``Waveglow: A flow-based generative
  network for speech synthesis,'' in \emph{ICASSP 2019-2019 IEEE International
  Conference on Acoustics, Speech and Signal Processing (ICASSP)}.\hskip 1em
  plus 0.5em minus 0.4em\relax IEEE, 2019, pp. 3617--3621.

\bibitem{oord2016wavenet}
A.~v.~d. Oord, S.~Dieleman, H.~Zen, K.~Simonyan, O.~Vinyals, A.~Graves,
  N.~Kalchbrenner, A.~Senior, and K.~Kavukcuoglu, ``Wavenet: A generative model
  for raw audio,'' \emph{arXiv preprint arXiv:1609.03499}, 2016.

\bibitem{takahashi2021source}
N.~Takahashi, M.~K. Singh, and Y.~Mitsufuji, ``Source mixing and separation
  robust audio steganography,'' \emph{arXiv preprint arXiv:2110.05054}, 2021.

\bibitem{dinh2014nice}
L.~Dinh, D.~Krueger, and Y.~Bengio, ``Nice: Non-linear independent components
  estimation,'' \emph{arXiv preprint arXiv:1410.8516}, 2014.

\bibitem{kingma2018glow}
D.~P. Kingma and P.~Dhariwal, ``Glow: Generative flow with invertible 1x1
  convolutions,'' \emph{Advances in neural information processing systems},
  vol.~31, 2018.

\bibitem{ardizzone2019guided}
L.~Ardizzone, C.~L{\"u}th, J.~Kruse, C.~Rother, and U.~K{\"o}the, ``Guided
  image generation with conditional invertible neural networks,'' \emph{arXiv
  preprint arXiv:1907.02392}, 2019.

\bibitem{xiao2020rescale}
M.~Xiao, S.~Zheng, C.~Liu, Y.~Wang, D.~He, G.~Ke, J.~Bian, Z.~Lin, and T.-Y.
  Liu, ``Invertible image rescaling,'' in \emph{Computer Vision--ECCV 2020:
  16th European Conference, Glasgow, UK, August 23--28, 2020, Proceedings, Part
  I 16}.\hskip 1em plus 0.5em minus 0.4em\relax Springer, 2020, pp. 126--144.

\bibitem{jing2021hinet}
J.~Jing, X.~Deng, M.~Xu, J.~Wang, and Z.~Guan, ``Hinet: deep image hiding by
  invertible network,'' in \emph{Proceedings of the IEEE/CVF international
  conference on computer vision}, 2021, pp. 4733--4742.

\bibitem{guan2022deepmih}
Z.~Guan, J.~Jing, X.~Deng, M.~Xu, L.~Jiang, Z.~Zhang, and Y.~Li, ``Deepmih:
  Deep invertible network for multiple image hiding,'' \emph{IEEE Transactions
  on Pattern Analysis and Machine Intelligence}, vol.~45, no.~1, pp. 372--390,
  2022.

\bibitem{ren2022color}
Y.~Ren, T.~Liu, L.~Zhai, and L.~Wang, ``Hiding data in colors: Secure and
  lossless deep image steganography via conditional invertible neural
  networks,'' \emph{CoRR}, vol. abs/2201.07444, 2022.

\bibitem{xu2022robust}
Y.~Xu, C.~Mou, Y.~Hu, J.~Xie, and J.~Zhang, ``Robust invertible image
  steganography,'' in \emph{Proceedings of the IEEE/CVF Conference on Computer
  Vision and Pattern Recognition}, 2022, pp. 7875--7884.

\bibitem{ma2022cin}
R.~Ma, M.~Guo, Y.~Hou, F.~Yang, Y.~Li, H.~Jia, and X.~Xie, ``Towards blind
  watermarking: Combining invertible and non-invertible mechanisms,'' in
  \emph{Proceedings of the 30th ACM International Conference on Multimedia},
  2022, pp. 1532--1542.

\bibitem{yamamoto2020parallel}
R.~Yamamoto, E.~Song, and J.-M. Kim, ``Parallel wavegan: A fast waveform
  generation model based on generative adversarial networks with
  multi-resolution spectrogram,'' in \emph{ICASSP 2020-2020 IEEE International
  Conference on Acoustics, Speech and Signal Processing (ICASSP)}.\hskip 1em
  plus 0.5em minus 0.4em\relax IEEE, 2020, pp. 6199--6203.

\bibitem{desplanques2020ecapa}
B.~Desplanques, J.~Thienpondt, and K.~Demuynck, ``Ecapa-tdnn: Emphasized
  channel attention, propagation and aggregation in tdnn based speaker
  verification,'' \emph{arXiv preprint arXiv:2005.07143}, 2020.

\bibitem{ke2024stegformer}
X.~Ke, H.~Wu, and W.~Guo, ``Stegformer: Rebuilding the glory of
  autoencoder-based steganography,'' in \emph{Proceedings of the AAAI
  Conference on Artificial Intelligence}, vol.~38, no.~3, 2024, pp. 2723--2731.

\end{thebibliography}

\vfill

\end{document}